\documentstyle[prb,aps,multicol]{revtex} \tighten
\begin{document}
\draft
\title{Negative electrostatic contribution to the bending rigidity of
charged
membranes and
polyelectrolytes screened by multivalent counterions}
\author{T. T. Nguyen, I. Rouzina and B. I. Shklovskii}
\address{Theoretical Physics
Institute, University of Minnesota, 116 Church St. Southeast,
Minneapolis, Minnesota 55455}
\maketitle

\begin{abstract}

Bending rigidity of a charged membrane or polyelectrolyte screened
by monovalent counterions is known to be enhanced by electrostatic
effects.
We show that in the case of screening by multivalent counterions
the electrostatic effects reduce the bending rigidity.
This inversion of the sign of the electrostatic contribution is related to 
the formation of two-dimensional strongly correlated liquids (SCL)
of counterions at the charged surface due to strong lateral repulsion
between them. 
When a membrane or a polyelectrolyte is bent, SCL is compressed 
on one side and stretched on the other
so that thermodynamic properties of SCL
contribute to the bending rigidity. Thermodynamic properties of SCL are 
similar to those of Wigner crystal and are anomalous in the sense that
the pressure, compressibility and screening radius of SCL are negative. 
This brings about substantial negative correction to the bending
rigidity.
For the case of DNA this effect qualitatively agrees with experiment.

\end{abstract}

 \begin{multicols}{2} 

\section {Introduction}

Many polymers and membranes are strongly charged in a water solution.
Among 
them are biopolymers
such as lipid membranes, DNA, actin and other proteins as well 
as numerous synthetic
polyelectrolytes. In this paper, we concentrate on bending of
membranes and cylindrical polyelectrolytes with fixed uniform 
distribution of charge at
 their surfaces. For a flat symmetrical membrane, the curvature
free energy per unit area can be expressed
in terms of the curvatures $c_1$ and $c_2$ along two orthogonal 
axes as~\cite{safran}
\begin{equation}
\frac{\delta F}{S}=\frac{1}{2}\kappa(c_1+c_2)^{2}
 + \kappa_{G} c_1 c_2 \label{definitions}
~~,
\end{equation}
where $\kappa$ is the bending rigidity, $\kappa_{G}$ 
is the Gaussian rigidity and
$S$ is the membrane surface area.
For cylindrical and spherical deformations with the radius of curvature
$R_{c}$
(see Fig.~1)
\begin{eqnarray}
\frac{\delta F^{cyl}}{S}&=&\frac{1}{2} \kappa R_{c}^{-2}, \label{cyl} \\
\frac{\delta F^{sph}}{S}&=&(2\kappa + \kappa_G )R_{c}^{-2} \label{spher}
\end{eqnarray}
respectively. In general, $\kappa=\kappa_{0}+\kappa_{el}$, where
$\kappa_{0}$ is the ``bare"
bending
rigidity related to short range forces and $\kappa_{el}$ is
electrostatic
contribution
which is determined by the magnitude of surface charge density and
the condition of
its screening by small ions of the water solution.
 Similarly, for a rod-like polymer,
such as double helix DNA, the change in free energy per unit length due
to
bending is given by
\begin{equation}
\frac{\delta F}{{\cal L}}=\frac{1}{2}Q R_{c}^{-2}, \label{rodF}
\end{equation}
where $\cal L$ is the length of the rod, $Q=Q_{0}+Q_{el}$ is the bending 
constant of the rod, which consist of
a "bare" component,
$Q_{0}$, and an electrostatic contribution $Q_{el}$. In the worm model
of
a linear polymer,
the persistence length, $L$, of the polymer is related to $Q$: 
\begin{equation}
L=\frac{Q}{k_{B}T}= \frac{Q_{0}}{k_{B}T} + \frac{Q_{el}}{k_{B}T} = L_{0}
+
L_{el},
\label{persistent}
\end{equation}
where $L_{0}$ is the bare persistent length and 
$L_{el}$ is an electrostatic contribution to it.
In the absence of screening, repulsion of like charges of a
membrane or a polyelectrolyte
leads to infinite
$\kappa_{el}$ and $L_{el}$. Only screening makes them finite. 
When the surface charge density is small 
enough Debye-H\"{u}ckel (DH) approximation can be used. 
For a membrane with the
surface charge density $-\sigma$ on each side, $\kappa_{el}$ was calculated~
\cite{winter,lekker,Duplantier,pincus}
when DH screening length $r_{s}$ is larger than membrane thickness $h$:
\begin{equation}
\kappa_{DH}=3\pi \frac{\sigma^{2}r_{s}^{3}}{D},
 ~~~~~\kappa_{G,DH} = - \frac{2}{3} \kappa_{DH}~~~~(h \ll r_s).
 \label{kappaDH}
\end{equation}
Here $D$ is dielectric constant of water. 

For cylindrical polyelectrolyte with diameter $d$ much smaller than
$r_{s}$,
calculations in the DH limit lead
to the well known Odijk-Skolnick-Fixman formula~\cite{odijk}
 for the persistence length:
\begin{equation}
L_{DH}=
\frac{\eta^{2}r_{s}^{2}}{4Dk_{B}T}~~~~~~(d \ll r_s),
 \label{LDH} \end{equation}
where $-\eta = \pi \sigma d$ is the charge per unit length of the
polymer.
Eqs. (\ref{kappaDH}) and
(\ref{LDH}) show that, in DH approximation, $\kappa_{el}$ 
and
$L_{el}$ vanish
at $r_s=0$ so that one can measure $\kappa_{0}$ 
and $L_{0}$ in the limit of high concentration of
monovalent salt.
 At at $r_s > 0$, the quantities $\kappa_{el}$ 
and $L_{el}$ are always positive and grow with $r_s$. 
However, in many practical situations,
polyelectrolytes are so strongly charged that DH approximation 
does not work and
the nonlinear Poisson-Boltzmann (PB) equation was used to 
calculate $\kappa_{el}$ and $L_{el}$.
If counterions have charge $Ze$, PB equation gives, for
 a thin membrane~\cite{lekker}
\begin{equation}
\kappa_{PB}=\frac{k_{B}T r_{s}}{\pi l}~~~~ \kappa_{G,PB} =
- \frac{\pi^{2}}{3}\kappa_{PB}~~~~(h \ll r_s)
 \label{kappaPB}
\end{equation}
and for the thin rod~\cite{LeBret}
\begin{equation}
L_{PB}=\frac{ r_{s}^{2}}{4l}~~~~~~(d \ll r_s), \label{LPB} 
\end{equation}
where $l=Z^{2}e^{2}/Dk_{B}T$ is the Bjerrum length with charge $Z$.
 Eqs.~(\ref{kappaDH}), (\ref{LDH}), 
(\ref{kappaPB}), and (\ref{LPB}) give positive $\kappa_{el}$ and
$L_{el}$ in agreement
 with the common expectations that electrostatic effects can only
increase bending rigidity. 

This paper deals with the case of a strongly charged membrane or 
polyelectrolyte with a uniform distribution of immobile charge on its 
surface. 
It was shown in
Ref.~\onlinecite{Gulbrand,Roland,Rouzina96,Bruinsma,Levin,Shklov98,Shklov99}
 that screening of such surface by multivalent counterions
with charge $Z \geq 2$
can not be described by PB equation. Due to strong 
lateral Coulomb repulsion,
counterions condensed on the surface form strongly correlated
two-dimensional liquid (SCL). Their correlations are so strong that a  
simple picture of the two-dimensional
Wigner crystal (WC) of counterions on a background of uniform 
surface charge
is a good approximation for calculation of the free energy of the SCL.
The concept of SCL was used to demonstrate that 
two charged surfaces in the 
presence of multivalent counterions attract each other at small
 distances~\cite{Rouzina96,Shklov98,Shklov99}.
It was also shown that cohesive energy
of SCL leads to much stronger counterion attraction to the surface
than in conventional
solutions of Poisson-Boltzmann equation, 
so that surface charge is almost totally
compensated by the SCL~\cite{Shklov99}. 

In this paper we calculate effect of SCL at the surface of
a membrane or a polyelectrolyte
on its bending rigidity. When a membrane or 
polyelectrolyte is bent, the density
of its SCL follows the changes in the density of the surface charge, 
increasing on one side and decreasing on the opposite side of
(see fig. 1).
As a result the bending  
rigidities can be expressed through thermodynamic properties of the SCL,
namely 
two-dimensional pressure and compressibility.
For two-dimensional one component plasma (on uniform background) these
 quantities were found
by Monte-Carlo simulation and other numerical 
methods~\cite{Totsuji,Lado,Gann} as
functions of temperature.
The inverse dimensionless temperature of SCL is usually written as the 
ratio of the average Coulomb 
interaction between ions to the thermal kinetic energy $k_BT$
\begin{equation}
\Gamma = \frac{(\pi n)^{1/2}Z^{2}e^{2}}{D k_BT},
\label{Gamma}
\end{equation}
where $n=\sigma/Ze$ is concentration of SCL. (For e.g., for $Z=3$ and
 $\sigma = 1.0~e/$nm$^{-2}$, $\Gamma = 6.3$).
We will show that in the range of our interest $3 < \Gamma <15$ 
the free energy, the pressure and
the compressibility and, therefore, electrostatic bending rigidities
differ 
only by $20\%$ from those 
in the low temperature limit $\Gamma \rightarrow \infty$, 
when SCL freezes into WC.
General results are given in Sec. III.
Here we present very simple results obtained in the WC limit:
\begin{eqnarray}
\kappa_{WC}&=& - 0.68 \frac{\sigma^{2}}{D}h^{2}a =
 - 0.74 \frac{\sigma^{3/2}(Ze)^{1/2}h^{2}}{D} ~~,\nonumber \\
 &&~~~ \kappa_{G,WC}=-\frac{2}{3}\kappa_{WC} ~~, \label{kappaWC} \\ 
L_{WC}&=& - 0.054\frac{\eta^{2}}{Dk_{B}T}da = 
- 0.10 \frac{\eta^{3/2}(Ze)^{1/2}d^{3/2}}{Dk_{B}T} ~~.\label{LWC} 
\end{eqnarray}
Here $a = (2Ze/\sqrt{3}\sigma)^{1/2}$ is the lattice constant of 
the triangular close packed WC. The 
membrane and the cylinder are assumed to be reasonably thick, 
$2\pi h \gg a$ and $\pi d \gg a$.
In contrast with results for DH
and PB approximations, $\kappa_{WC}$ and $L_{WC}$ are $negative$, 
so that multivalent
counterions make a membrane or a polyelectrolyte more flexible.
 For a membrane with $\sigma = 1.0~e/$nm$^{-2}$, $h=4$ nm
at $Z=3$ we find that $a = 1.7$ nm, inequality  
$2\pi h \gg a$ is fulfilled and
 Eq.~(\ref{kappaWC}) yields $\kappa_{WC} = - 14 k_{B}T$
 (at room temperature). This value should be compared with typical 
$\kappa_{0} \sim 20-100 k_{B}T$. 
For a cylindrical polyelectrolyte with parameters of the double
helix DNA, $d =2~$nm and $\eta = 5.9~e/$nm, inequality $\pi d \gg a$
is valid and we obtain $L_{WC} = - 4.9$ nm, which is much smaller
than the bare persistence length $L_0 = 50$ nm. We should, however,
note that our estimates are based on the use of the bulk dielectric
constant of water $D=80$. For the lateral interactions of counterions
near  the surface of organic material with low dielectric constant, the
effective $D$ can be substantially smaller. 
(In macroscopic approach it is close to $D/2$). 
As a result, absolute values of 
$\kappa_{WC}$ and $L_{WC}$ can grow significantly.

Negative electrostatic contributions to the bending 
rigidity were also predicted
 in two recent papers~\cite{Pincus98,Kardar}.
The authors considered this problem in the high temperature limit 
where attraction between different points of a membrane or a 
polyelectrolyte is a result of correlations 
of thermal fluctuations of screening atmosphere at these points.
 Such theories describe negative contribution to rigidity for 
$Z=1$ or for larger Z but with weakly charged surfaces where $\Gamma < 1$.
On the other hand, at $Z \geq 3$ and large $\sigma$, one deals with
low temperature situation when $\Gamma \gg 1$.
In this case the main terms of the electrostatic 
contribution to the bending rigidity
are given by Eq.~(\ref{kappaWC}) and Eq.~(\ref{LWC}), which are based on 
$static$ spatial correlations of ions.


We would like to emphasize that, contrary to Ref.~\onlinecite{Kardar},
this paper deals only with small deformations
of a membrane or a polyelectrolyte.  We are 
not talking about a global instability of a membrane or 
 polyelectrolyte due 
to self-attraction,
where, for example, a membrane rolls itself into a cylinder 
or a polyelectrolyte, as in the case
of DNA, rolls into a toroidal particle~\cite{Rouzina96}.
 Global instabilities
 can happen even when total
local bending rigidities are still positive. To prevent
 these instabilities in experiment one can work with a
small area membrane or short polyelectrolyte~\cite{Porchke} or 
keep their total bend small by an external force, for example, with 
optical tweezers~\cite{Baumann,Rouzina98}.
 
It is  known that, in a monovalent salt, DNA has a persistence length 
$L > 50$ nm which saturates at 50 nm at 
large concentration of salt. Thus it is natural to assume that
the bare persistence length $L_0 = 50$ nm. 
However, it was found in Ref.~\onlinecite{Porchke,Baumann,Rouzina98}
that a relatively small
concentration of counterions with $Z=2,3,4$ leads 
to an even smaller persistence length, which can be
as low as $L = 25-30$ nm. We emphasize that a strong effect
was observed for multivalent counterions which are known to bind to DNA
due to the non-specific electrostatic force. 

These experimental data can be interpreted as a result of
replacement of monovalent counterions with multivalent 
ones which create SCL at the DNA surface.
As we stated before, multivalent counterions should 
produce a negative correction to $L_0$, although the above calculated  
correction to persistence length is smaller than the experimental
one. 

This paper is organized as follows. In Sec. II we discuss
thermodynamic properties
of SCL and WC as functions of its density and temperature. 
In sec. III and IV we use
expressions for their pressure and 
compressibility to calculate $\kappa_{SCL}$ and $L_{SCL}$
and their asymptotic expressions
$\kappa_{WC}$ and $L_{WC}$.
In Sec. V we calculate contributions of the tail of screening atmosphere
to
$\kappa_{el}$ and $L_{el}$ and show that for $Z \geq 2$ and strongly
charged 
membranes and polyelectrolytes, tail contributions to  the bending
rigidity 
are small in comparison with that of SCL.

\section{Strongly correlated liquid and Wigner Crystal} 

Let us consider a flat surface uniformly charged with surface density 
$-\sigma$ and covered by
concentration $n = \sigma/ Ze$ of counterions with charge $Ze$. 
It is well known that the minimum of 
Coulomb energy of counterion repulsion and
their attraction
to the background is provided by a triangular close packed WC of
counterions.
Let us write energy per unit surface area of WC as $E=n \varepsilon(n)$
where $\varepsilon(n)$ is the energy per ion. One can estimate
 $\varepsilon(n)$ as the interaction
energy of an ion with its Wigner-Seitz cell of background charge 
(a hexagon of the background with charge $-Ze$). This estimate gives
$\varepsilon(n)\sim -Z^{2}e^{2}/Da \sim -Z^{2}e^{2}n^{1/2}/D$.
 More accurate expression for $\varepsilon(n)$ is~\cite{mara} 

\begin{equation}
\varepsilon(n)= - \alpha n^{1/2}Z^{2}e^{2}D^{-1} = -1.1\Gamma k_BT,
\label{energyion}
\end{equation}
where $\alpha=1.96$. At room temperature, 
Eq.~(\ref{energyion}) can be rewritten as 
\begin{equation}
\varepsilon(n)\simeq - 1.4~Z^{3/2}(\sigma/e)^{1/2}k_BT ~~, 
\label{estimate}
\end{equation}
\noindent where $\sigma/e$ is measured in units of nm$^{-2}$.

At $\sigma = 1.0~e/$nm$^{-2}$, Eq.~(\ref{estimate}) 
gives $|\varepsilon(n)|
\simeq 7 k_BT$ or $\Gamma = 6.3$ at $Z=3$, and $|\varepsilon(n)| 
\simeq 13 k_BT$ or  $\Gamma = 12$ at $Z=4$. Thus for
multivalent ions at room temperature we are dealing with the 
low temperature regime. However,
it is known \cite{Gann} that due to a very small shear modulus, WC melts
at even lower temperature: $\Gamma \simeq 130$.
 Nevertheless, the disappearance of
long range order produces only a small effect on thermodynamic
 properties. They are determined by
 the short range order which does not change significantly in 
the range of our interest
 $5 < \Gamma < 15$~\cite{Rouzina96,Bruinsma,Shklov98,Shklov99}. 
This can be seen from
 numerical calculations~\cite{Totsuji,Lado,Gann} of thermodynamic 
properties of classical 
two-dimensional SCL of Coulomb particles on the neutralizing
background. In the range $0.5 < \Gamma < 50$, the internal
 energy of SCL per counterion,
 $\varepsilon(n,T)$, was fitted by 
\begin{equation}
\varepsilon(n,T) = k_BT ( - 1.1 \Gamma + 0.58 \Gamma^{1/4} + 0.74),
 \label{intenergy}
\end{equation}
with an error less than 2\% ~\cite{Totsuji}. The first term on the right
side of
Eq. (\ref{intenergy})
is identical to Eq.~(\ref{energyion}) and dominates at large $\Gamma$. 
All other thermodynamic
functions can be obtained from Eq.~(\ref{intenergy}).
In the next section we show that $\kappa_{el}$ and $L_{el}$ are
proportional
to the
inverse isothermal compressibility of SCL at a given number of ions
$N$
\begin{equation}
\chi^{-1}= n(\partial P/\partial n)_T,
\label{compressdef}
\end{equation}
where
\begin{eqnarray}
P &=& - (\partial F/\partial S)_T = (n\varepsilon(n,T) + n
k_BT)/2\nonumber \\
&=& nk_BT ( - 0.55 \Gamma + 0.27 \Gamma^{1/4} + 0.87)
\label{pressure}
\end{eqnarray}
is the two-dimensional pressure,
$F$ is the free energy of SCL and $S=N/n$ is its area.
 Using Eq.~(\ref{pressure}) and relation $\partial\Gamma/\partial n 
= \Gamma/2n$, one finds
\begin{equation}
\chi^{-1}= nk_BT ( - 0.83 \Gamma + 0.33 \Gamma^{1/4} + 0.87), 
\label{compress}
\end{equation}
where the first term on the right side follows from
Eq.~(\ref{energyion})
 and describes WC limit. The last two terms give 33\% 
correction to the WC term at $\Gamma = 5$ and only 12\% correction at
$\Gamma = 15$.
So one can use zero temperature, Eq.~(\ref{energyion}),
as first approximation 
to calculate $\kappa_{el}$ and $L_{el}$. This is how we obtained 
Eq.~(\ref{kappaWC}) and Eq.~(\ref{LWC}). 

Eqs.~(\ref{pressure}) and (\ref{compress}) show that, in contrast with
most of
liquids and solids, SCL and WC have $negative$ pressure $P$ and 
compressibility $\chi$. 
We will see below that anomalous behavior is the reason
 for anomalous $negative$ rigidity $\kappa_{el}$ and persistence length
$L_{el}$
and $positive$ Gaussian rigidity $\kappa_{G,el}$. The curious negative
sign
 of compressibility of two-dimensional electron SCL and WC was first 
predicted in Ref.~\onlinecite{Bello}. Later it was discovered in 
magneto-capacitance experiments in MOSFETs 
and semiconductor heterojunctions~\cite{Krav,Eisen}. 

According to Eq.~(\ref{compress})
$\chi^{-1} = 0$ at $\Gamma = 1.48$, $P=0$ at $\Gamma=2.18$ 
and they become positive at smaller
$\Gamma$.
As one can see from Eqs.~(\ref{estimate}) and (\ref{Gamma}),
 at $\sigma \sim 1.0~e/$nm$^{-2}$
such small values of $\Gamma$ correspond to $Z=1$. Thus 
surface layer of monovalent ions do not produce large 
negative $\kappa_{el}$ and $L_{el}$ in comparison with multivalent ions. 
For them conventional results of Eqs.~(\ref{kappaDH}), (\ref{LDH}), 
(\ref{kappaPB}),
and (\ref{LPB}) related with counterions in the long distance tail 
of screening atmosphere work better.
We will return to this question in Sec. V where we discuss the role of
these
tails.

\section {Membrane}

We will consider a ``thick" membrane for which one can neglect 
the effects of the correlation of SCL on two surfaces  of the membrane.
If we approximate SCL by WC, the energy of 
such correlations between two surfaces of the membrane decay as
 $\exp(-2\pi h/a)$, so the condition of ``thickness", 
$h \gg 2\pi a$, is actually easily satisfied 
for a strongly charged membrane.

Let us first write the free energy of each surface of the membrane as 
\begin{equation}
F=Nf(n,T)
\end{equation}
where $f(n,T)$ is the free energy per ion. 
\begin{figure}[h]
\input{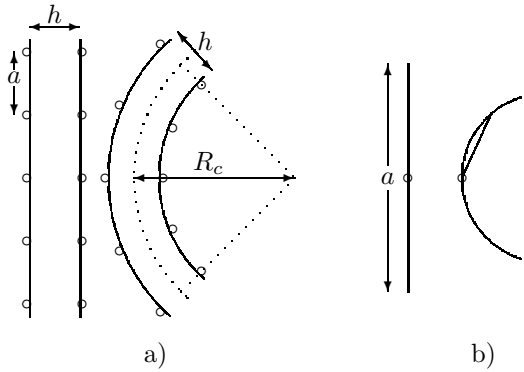}
\caption{Bending of membrane (the curvature has been exaggerated).
 For simplicity, the WC case is depicted. a) A thick membrane.
The right WC is compressed while the left WC is stretched. 
For thick membranes, 
this is the dominant cause of the change in free energy. b) 
A very thin membrane. Only one Wigner-Seitz cell is shown.
Due to finite curvature of the surface,
the distance from any point of the Wigner-Seitz cell to the central 
ion is shorter than that in the flat configuration.
 For thin membranes, this is the dominant cause of free energy change.}
\end{figure}
When a membrane is bent (see Fig. 1a), the surface
 charge on the right side is compressed to a new 
density $n_{R} > n$, while the surface charge on 
the left side is stretched to $n_{L} < n$. Since 
the total charge on each surface is conserved, 
this change in density leads to a change in the 
free energy of each surface: 
\begin{equation}
\delta F_{L,R} = N\left(\frac{\partial f}{\partial n}\delta n_{L,R} + 
\frac{1}{2}\frac{\partial^{2} f}{\partial n^{2}}
\delta n_{L,R}^{2}\right) 
~~,
\label{dfn}
\end{equation}
\noindent in which we kept only terms up to second order 
in $\delta n_{L,R}=n_{L,R}-n$.

Using the definitions (\ref{pressure}) and (\ref{compressdef}) 
for the pressure and the compressibility of 2D systems 

\begin{eqnarray}
P&=&-\left(\frac{\partial F}{\partial S}\right)_{N,T} 
=-N\left(\frac{\partial f}{\partial S}\right)_{N,T}
 =n^{2}\frac{\partial f}{\partial n} ~~,\\
 \frac{1}{\chi}&=&n\left(\frac{\partial P}{\partial n}\right)_{T} 
= 2n^{2}\frac{\partial f}{\partial n}+
n^{3}\frac{\partial^{2} f}{\partial n^{2}}~~, \end{eqnarray}

\noindent Eq.~(\ref{dfn}) can be rewritten as 

\begin{equation}
\delta F_{L,R}=\frac{SP}{n}~\delta n_{L,R} + 
\frac{S}{n^2}(\frac{1}{2\chi} - P)~\delta n_{L,R}^2 ~~~. \end{equation}

So, the total change in the free energy of the membrane per unit area is

\begin{eqnarray}
\frac{\delta F}{S}&=&\frac{\delta F_{L}+\delta F_{R}}{S} 
=\frac{P}{n}(n_{L}+n_{R}-2n) + 
\nonumber \\ & & ~~~~\frac{1}{n^2}(\frac{1}{2\chi}
 - P)((n_{L}-n)^{2}+(n_{R}-n)^{2}) ~~.
\label{dfLR}
\end{eqnarray}

In the case of cylindrical geometry,
keeping only terms up to second order in the curvature $R_{c}^{-1}$, we
have
\begin{equation}
n_{L,R}=\frac{R_{c}}{R_{c}\pm h/2} n
\simeq \left(1\mp\frac{h}{2R_{c}}+\frac{h^{2}}{4R_{c}^{2}}\right)n ~~.
\label{nLR}
\end{equation}
Substituting Eq.~(\ref{nLR}) into Eq.~(\ref{dfLR}), we get 
\begin{equation}
\frac{\delta F^{\rm cyl}}{S}
=\frac{1}{4\chi}h^{2}R_{c}^{-2} ~~.
\label{fcyl}
\end{equation}

Similarly, in the case of spherical geometry we have 

\begin{equation}
n_{L,R}=\left(\frac{R_{c}}{R_{c}\pm h/2}\right)^{2} n \simeq 
\left(1\mp\frac{h}{R_{c}}+\frac{3h^{2}}{4R_{c}^{2}}\right)n 
\end{equation}
\noindent and
\begin{eqnarray}
\frac{\delta F^{\rm sphere}}{S}
=\left(\frac{1}{\chi}-\frac{P}{2}\right)h^{2}R_{c}^{-2} ~~.
\label{fsphere}
\end{eqnarray}

Comparing Eq.~(\ref{fcyl}) and (\ref{fsphere}) with Eq.~(\ref{cyl}) 
and (\ref{spher}), 
we obtain general expressions for the electrostatic contribution to the
bending rigidity
\begin{equation}
\kappa_{el} = \frac{h^2}{2\chi}\: , ~~~~ \kappa_{G,el}= -\frac{h^ 2
P}{2}
 \label{kappael} ~~.
\end{equation}
For example, in the case of low surface charge density, 
DH approximation can be used to get~\cite{winter}

\begin{equation}
f(n,T)=2\pi \frac{\sigma^2}{D} n^{-1}r_{s} ~~, \end{equation}

\noindent from which, we can easily get a generalization of 
Eq. (\ref{kappaDH}) for a ``thick" membrane ($h \gg r_{s}$) 

\begin{equation}
\kappa_{DH}=2\pi\frac{\sigma^{2}}{D}h^{2}r_{s}, 
~~~~ \kappa_{G,DH}=-\frac{1}{2}\kappa_{DH}
~~.
\label{kappaDHthick}
\end{equation}

In the case of high surface charge density we study in this paper, a SCL 
of multivalent counterions resides on each surface of the membrane.
The expressions for the pressure and the compressibility given by
 Eqs.~(\ref{pressure}) and (\ref{compress}) can be used to calculate 
the bending rigidity:

\begin{eqnarray}
\kappa_{SCL}&=&\frac{nh^{2}}{2}
k_BT(-0.83 \Gamma + 0.33 \Gamma^{1/4} + 0.87) ~~,
\label{kappaSCLthick} \\
\kappa_{G,SCL}&=&-\frac{nh^{2}}{2}
k_BT(-0.55 \Gamma + 0.27 \Gamma^{1/4} + 0.87) ~~.
\label{kappaGSCLthick}
\end{eqnarray}

In the limit of a strongly charged surface ($\Gamma \gg 1$), 
the first term in 
Eqs.~(\ref{kappaSCLthick}) and (\ref{kappaGSCLthick})
 dominates, the free energy
of SCL is close to that of WC.
Using Eq.~(\ref{Gamma}) one arrives at Eq.~(\ref{kappaWC}) 
for the bending rigidity in the WC limit. 

As already stated in Sec. 1, for $\Gamma > 3$,
Eqs.~(\ref{kappaSCLthick}),
 (\ref{kappaGSCLthick}) give a negative value for the 
bending modulus and a positive value 
for the Gaussian bending modulus. In other words, multivalent
 counterions make the membrane more flexible. 
This conclusion is opposite to the standard results
 obtained by mean field theories 
(Eqs.~(\ref{kappaDH}), (\ref{kappaPB}), (\ref{kappaDHthick})) where
electrostatic effects are known to enhance the bending rigidity of 
membranes ($\kappa_{el} > 0$
 and $\kappa_{G,el} < 0$). Obviously, this anomaly is related
 to the strong correlation between
 multivalent counterions condensed on the surface of the membrane,
 which was neglected in mean field theories.

We can also look at Eqs.~(\ref{kappaDHthick}) 
and (\ref{kappaWC}) from 
another interesting perspective:
 apart from a numerical factor,
 Eq.~(\ref{kappaDHthick}) is identical to Eq.~(\ref{kappaWC}) if 
 we replace $r_{s}$ by $-a$.
 So the WC of counterions has effect on bending properties of the
 membrane 
 as if one replaces the
 normal 3D screening length of counterions 
 gas by a $negative$ screening length 
 of the order of lattice constant.
 Such negative screening length of WC or SCL has been derived 
 for the first time  in
 Ref.~\onlinecite{Efros}. It follows from the negative
 compressibility predicted in Ref.~\onlinecite{Bello}, 
 and observed in Refs.~\onlinecite{Krav} and \onlinecite{Eisen}.

Until now we have ignored the effects related to
Poisson's ratio $\sigma_{P}$ of the membrane material.
We are talking about the
bending induced increase of the thickness 
of the compressed (right) half of the membrane, simultaneous decrease
of the thickness of its stretched (left) half, and 
the corresponding shift
of the neutral plane of the membrane 
(the plane which by definition does not 
experience any compression or stretching) 
to the left from the central plane.
These deformations can be found following Ref.~\onlinecite{Landau}
and lead to additional term 
$\sigma_{P}h^2/(1 - \sigma_{P})R_{c}^2$ in the right side of
Eq.~(\ref{nLR}). It gives for the bending rigidity

\begin{equation}
\kappa_{el} = \frac{h^2}{2\chi} + 
\frac{\sigma_P}{1 - \sigma_P}\frac{Ph^2}{2}.
\label{kappaP}
\end{equation}
So, for example, at $\sigma_P =1/3$, the second term of  
Eq.~(\ref{kappaP}) gives a 33\% correction to
Eq.~(\ref{kappaWC}).

According to Eqs.~(\ref{kappael}), (\ref{kappaSCLthick}),
(\ref{kappaGSCLthick}) $\kappa_{el} = 0$ at $h=0$.
This happens because in this limit two SCL
 merge into one, whose surface charge density
remains unchanged after bending. Nevertheless,
there is another effect directly related to the
 curvature of SCL. It can be explained 
by concentrating on one curved Wigner-Seitz cell (see Fig. 1b).
 One can see, that due to the curvature,
 points of the background come closer to the central
 counterion of the cell 
in the three-dimensional space where Coulomb interaction operates.
 As a result, the energy of SCL goes down.  
In the Wigner-Seitz approximation, where 
energy per ion of WC is approximated by its interaction with 
the Wigner-Seitz cell of the background charge, we obtain
\begin{equation}
\kappa_{WC}^{thin} \simeq -0.006\frac{\sigma^{2}a^{3}}{D} , 
~~~~\kappa_{G,WC}^{thin}=-\frac{2}{3}\kappa_{WC}^{thin} ~~~.
\label{kappaWCthin}
\end{equation}
We see that this effect also gives anomalous signs for 
electrostatic contribution to rigidity
in the WC limit, but with a very small numerical coefficient. 
Also note that, as in the thick membrane case, 
we can obtain Eq.~(\ref{kappaWCthin}) for a thin membrane by 
replacing $r_s$ in Eq.~(\ref{kappaDH}) by
a negative screening radius of WC with absolute value of the order $a$.

\section {Cylindrical polyelectrolytes}

In this section, we study bending properties of cylindrical
 polyelectrolytes with diameter 
$d$ and linear charge density $\eta$ (see Fig. 2).
As in the membrane problem, we will assume that the cylinder 
is thick, i.e. its circumference 
$\pi d$ is much larger than the average distance $a$ between 
counterions on it surface. 
The calculation is carried out exactly in the same way as in 
the case of thick membrane. The only
difference is that, instead of summing the free energy of two 
surfaces of the membrane, we
average over the circumference of the cylinder. 

Let us denote by $n_{\phi}$ the local density at an angle 
$\phi$ on the circumference
on the cylinder (see Fig. 2a). Before bending 
$n_{\phi}=n=\eta/\pi dZe$, after bending it changes to a new value
\begin{eqnarray}
n_{\phi}&=&n\frac{R_{c}}{R_{c}-(d/2)\cos\phi } \nonumber \\
 &\simeq&n\left(1+\frac{d\cos\phi}{2R_c}+ 
\frac{d^2\cos^{2}\phi}{4R_c^2} \right) ~~.
\label{nphi}
\end{eqnarray}
Using Eq.~(\ref{dfLR}) the free energy per unit length of
 the polymer can be written as

\begin{eqnarray}
\frac{\delta F}{\cal L}&=&\int_{0}^{2\pi} \frac{d}{2} ~d\phi 
\left(\frac{P}{n}(n_{\phi}-n) + \frac{1}{n^2}(\frac{1}{2\chi}-P)
 (n_{\phi}-n)^{2}\right) \nonumber \\
&=&\frac{\pi}{2\chi}\left(\frac{d}{2}\right)^{3}R_{c}^{-2} \label{dfphi}
\end{eqnarray}

\noindent where we keep terms
up to second order in the curvature $R_{c}^{-1}$.

\begin{figure}[h]
\begin{center}
\input{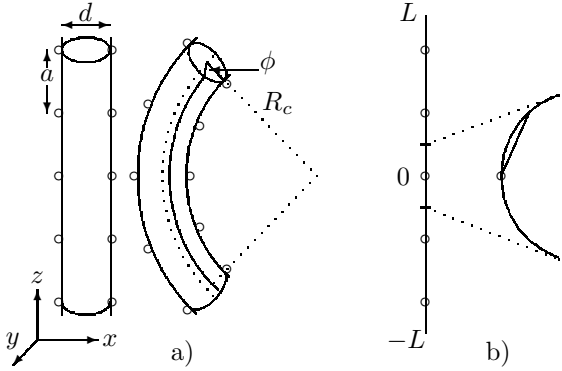}
\caption{Bending of cylindrical polyelectrolytes. 
a) A thick cylinder. Rigidity is mostly determined by the change
 in density of SCL.
b) A thin cylinder. The curvature effect, is the dominant cause of
change in free energy. }
\end{center}
\end{figure}

Comparing Eq.~(\ref{dfphi}) with Eq.~(\ref{rodF}),
(\ref{persistent}),
one can easily calculate the electrostatic
contribution to the persistence length
\begin{equation}
L_{el}=\frac{\pi}{\chi k_BT}\left(\frac{d}{2}\right)^{3} ~~.
\end{equation}
In the case of highly charged polymer, a SCL of counterions
 resides on the polymer surface.
For a thick cylinder, the SCL is locally flat
and we can use the numerical
expression (\ref{compress}) for $\chi^{-1}$ to obtain

\begin{equation}
L_{SCL}=\frac{\pi}{8}nd^{3}
( - 0.83 \Gamma + 0.33 \Gamma^{1/4} + 0.87) ~~.
\label{LSCLthick}
\end{equation}

Again, we see that correlations between counterions on the 
surface of a polymer lead
to a negative electric contribution to 
persistence length for $\Gamma > 1.5$.
In the WC limit $\Gamma \gg 1$, the first term in
 Eq.~(\ref{LSCLthick}) dominates, and 
using Eq.~(\ref{Gamma}) one can easily obtain Eq.~(\ref{LWC}).

As in the membrane case, for simplicity, in writing down Eqs. (\ref{nphi}),
we have ignored the effect of
finite value of the Poisson's ratio of the polymer material.
In membranes, this effect result in a gain in energy
due to the shift of the neutral plane toward the convex (stretched) sides.
For a cylinder, there is an additional expansion in the $y$ 
direction (Fig. 2) which reduces the change in surface charge density,
hence compensates the above gain. 
These deformations can be found following Ref. \onlinecite{Landau}
and lead to a correction to Eqs. (\ref{nphi})

\begin{eqnarray}
n_{\phi}&=&n\left(1+\frac{d\cos\phi}{2R_c}(1-\sigma_P)+
\frac{d^2\cos^2\phi}{4R_c^2}(1-\frac{\sigma_P}{2}+\sigma_P^2)
\right. \nonumber \\
&&~~~~~~\left. -
\frac{d^2\sigma_P^2}{8R_c^2}(1-\cos^2\phi)\right)
~~.
\end{eqnarray}

This gives, for the persistence length,

\begin{equation}
L_{el}=\frac{\pi}{k_B T}\left(\frac{d}{2}\right)^3
\left(\frac{1}{\chi}(1-\sigma_P)^2+P(3\sigma_P-\sigma_P^2)\right)
~.
\label{Lp}
\end{equation}

Obviously, due to the expansion in $y$ direction, the correction to
energy is not as strong as in the membrane case. 
For example, at $\sigma_P=1/3$,
Eq. (\ref{Lp}) gives only 3\% correction to Eq. (\ref{LWC}).

According to Eqs.~(\ref{LSCLthick}) and (\ref{LWC}),
at $d=0$, $\kappa_{el}$ vanishes. In this limit,
 we have to directly include
 the curvature effect on one dimensional SCL as shown in Fig. 2b.
 As already
 mentioned in the previous section, after bending,
points on a Wigner-Seitz cell come closer to the central ion, which 
lowers the energy of the system. This effect can be calculated
easily in the WC limit. Let's consider the electron at the origin,
its energy can be written as
\begin{equation}
\varepsilon=\sum_i
\frac{Z^2e^2}{Dr_{i}}-\int_{-L}^{L}ds\frac{Ze\eta}{Ds} ~~,
\label{energy}
\end{equation}
\noindent where $r_i=ia$ and $s$ is the contour distance from our ion
to an lattice point $i$ and the element $ds$ of the background charge.
In the straight rod configuration the space distant is the same as the
contour distance, however after bending they change to

\begin{equation}
r_i^{\prime}\simeq r_i(1-r_i^2/24R_c^2)  ~~, ~~~  
s^{\prime}\simeq s(1-s^2/24R_c^2) ~~.
\label{newrs}
\end{equation}
Using these new distances to calculate the energy of
the bent rod and subtract Eq. (\ref{energy}) from it,
one can easily calculate the change in energy due to curvature and
the corresponding contribution to persistence length: 

\begin{equation}
L_{WC}^{thin}=-\frac{l}{96} ~~,
\end{equation}
which is negative and very small.
For e.g., for $Z=3,~4$, $L_{WC}^{thin}=-0.065$ nm and $-0.116$
 nm respectively.

\section{Contributions of the tail of the screening atmosphere}

In previous sections, we calculated the contribution of a SCL of 
counterions condensed on the surface
of a membrane or polyelectrolyte to their bending rigidity.
We assumed that charge
density $\sigma$ is totally compensated by  
the concentration $n= \sigma/Ze$. Actually, for example, for a membrane,
some concentration, $N(x)$, of counterions is distributed
at a distance $x$ from the surface in the bulk of solution 
(we call it the tail of the screening atmosphere). 

 The standard solution of PB equation
 for concentration $N(x)$ at 
$N(\infty)=0$ has a form
\begin{equation}
N(x) = {1\over{2 \pi l}}{1\over {(\lambda + x)^2}},
\label{GC}
\end{equation}
where $\lambda = Ze/(2 \pi l \sigma)$ is Gouy-Chapman length. 
At $ \Gamma \gg 1$, correlations in SCL provide additional 
strong binding for counterions,
which dramatically change the form of $N(x)$~\cite{Shklov99}. 
It decays exponentially at $\lambda \ll x \ll l/4$,
and at $x \gg l/4$ it behaves as
\begin{equation}
N(x) = {1\over{2 \pi l}}{1\over {(\Lambda + x)^2}}.
\label{GCnew}
\end{equation}
Here $\Lambda = Ze/(2 \pi l \sigma^*)$ is 
an exponentially large length and
$\sigma^*$ is the exponentially small uncompensated 
surface charge density at the distance $\sim l/4$.
In any realistic situation when $N(\infty)$ is finite or 
a monovalent salt is added to the solution,
Eqs.~(\ref{GC}) and (\ref{GCnew}) should be 
truncated at the screening radius $r_s$. Then 
the solution of the standard PB equation gives~\cite{lekker} 
Eq.~(\ref{kappaPB}) at $r_s \gg \lambda$ or Eq.~(\ref{kappaDH})
 at $r_s \ll \lambda$.
In the case of SCL, for realistic values of $r_s$ in the range 
$l/4 < r_s \ll \Lambda$, we obtain a contribution of 
the tail similar to Eq.~(\ref{kappaDH}) 
\begin{equation}
\kappa_{t}=3\pi \frac{(\sigma^*)^{2}r_{s}^{3}}{D}.
\label{kappaelt}
\end{equation}
At reasonable values of $r_s$, 
this expression is much smaller than $\kappa_{WC}$
due to very small values of the ratio $\sigma^*/\sigma$.

Now we switch to a cylindrical polyelectrolyte. 
In this case, the solution of the PB equation is known~\cite{Zimm} 
to confirm the main features of the                              
Onsager-Manning~\cite{Manning}                                                
picture of the counterion condensation. This solution depends
on              
relation between $|\eta|$ and $\eta_c= Ze/l$. In the case 
interesting for us,
 $|\eta| \gg \eta_c$, the counterion charge $|\eta| - \eta_c$
is                                                    
localized at the cylinder surface, while the charge           
$\eta_c$, is spread in the bulk of the
solution.                               
This means that at large distances the apparent charge
density                  
of the cylinder, $\eta_a$, equals $- \eta_c$ and does not depend
on             
$\eta$. Eq.~(\ref{LPB}) can actually be obtained from
Eq.~(\ref{LDH}) by substituting  $\eta_c$ for $\eta$.

It is shown in Ref.~\onlinecite{Shklov99} that 
at $\Gamma \gg 1$, the existence of SCL at the surface of the cylinder 
leads to substantial corrections to the Onsager-Manning theory.
Due to additional binding of counterions by SCL 
$|\eta_a| < |\eta_c|$ and is given by the expression
\begin{equation}
\eta_a = - \eta_c {{\ln [N(0)/N(\infty)]}\over{\ln(4 r_{s} / l)}}
~~,
\label{etaapp}
\end{equation}
where $N(0)$ is exponentially small concentration
at the distance $r \geq l/4$
from the cylinder axis, used in Ref.\onlinecite{Shklov99}
as a boundary condition for PB equation at $x=0$.
Therefore, one can  obtain for 
the tail contribution, the estimate from the above
using Eq.~(\ref{LPB}). For $Z=3$ and $r_s = 5$ nm this
gives $L_t < 1$ nm. For DNA, 
this contribution is much smaller than $L_{SCL} \simeq - 5$ nm.

\section{Conclusion}
We would like to conclude with the discussion of
approximations used in this study.
First, we assumed that the surface charges are
immobile. This is true for rigid polyelectrolytes, such as
double helical DNA or actin, as well as for frozen or
tethered membranes. But if the membrane is fluid, its charged
polar heads can move along the surface. In this case surface charges
can accumulate near a $Z$-valent counterion and screen it. Such
screening creates short dipoles oriented perpendicular to the
surface. Interaction energy between these dipoles is 
much weaker than the 
correlation energy of SCL. Therefore it produces negligible 
contribution
to the membrane rigidity. The mobility of the charged polar heads
eliminates effects of counterion correlation only 
in the situation where
the membrane has polar heads of two different charges, for example,
neutral and negative ones. In such a membrane, the
local surface charge density can grow due to the 
increase of local concentration
of negative heads. But if all of the closely packed
polar heads are equally charged their motion does not lead to
redistribution of the surface charges. Then our theory is valid again.

Another approximation which we used is that 
the surface charge is uniformly smeared. This can not be exactly true
because localized charges are always discrete. Nevertheless our approximation
makes sense if the surface charges are distributed evenly, and their
absolute value is much smaller than the counterion charge. For example,
when the surface charged heads have charge -$e$ and the counterion
charge is $Ze\gg e$, then the repulsion between counterions is
much stronger than their pinning by the surface charges. At
$Z\geq 3$ we seem to be close to this picture.
On the other hand, if the surface charges 
were clustered, for example, they form compact triplets, the trivalent
counterion would simply neutralize such cluster, creating a small
dipole. Obviously our theory would over-estimate electrostatic 
contribution to the bending rigidity in this case. 

All calculations in this paper were done for point like counterions.
Actually counterions have a finite size and one can wonder 
how this affects our results. Our results, of course,
make sense only if 
the counterion diameter is smaller than the average distance between them 
in SCL. For a typical surface charge density, $\sigma = 1.0~e/$nm$^{-2}$,
the average distance 
between trivalent ions is 1.7 nm, so that 
this condition is easily satisfied. The most important correction to
the energy is related to the fact that 
due to ion's finite size, the plane of the center of the counterion
charge can be located at some distance from 
the plane of location of the surface
charge. This creates an additional planar capacitor at each surface 
and results in a positive contribution to 
the bending rigidity similar to Eq.~(\ref{kappaDHthick})
which can compensate our negative 
contribution.
On the other hand, if the negative ions stick out of
the surface and the centers of counterions are in the same plane with 
centers of negative charge this effect disappears.

In general case, one can look at this problem from another angle.
Let us assume that the bare quantities $\kappa_0$ and $L_0$
are constructively defined as experimental values obtained 
in the limit of a high concentration of monovalent 
counterions. Let us also assume that the distances of closest approach of
monovalent and $Z$-valent counterions to the surface are 
the same. This means that the planar capacitor effect 
discussed above is already included
in the bare quantities $\kappa_0$ and $L_0$. Then the replacement 
of monovalent counterions by $Z$-valent will always lead to 
Eq.~(\ref{kappaWC}) and Eq.~(\ref{LWC}).

In summary, we have shown that condensation of multivalent
counterions on the surface of a charged membrane or polyelectrolyte
happens in the form of a strongly correlated Coulomb liquid, which
closely resembles a Wigner crystal.  Anomalous properties of
this liquid lead to the observable decrease of  the bending rigidity of
a membrane and polyelectrolyte.

\acknowledgements

We are grateful to V. A. Bloomfield and A. Yu. Grosberg 
for valuable discussions.
This work was supported by NSF DMR-9616880 (T. N. and  B. S.) and
NIH GM 28093 (I. R.).

\end{multicols}
\end{document}